\documentclass[reprint,twocolumn,aps,
showkeys,
prl,
notitlepage]{revtex4-2}

\usepackage{
amsmath,
amssymb,
graphicx,subfigure,caption,
comment,
}
\usepackage{bm}
\usepackage{color}  
\usepackage{extarrows}
\usepackage{appendix}
\usepackage{paralist}
\usepackage{titlesec}

\usepackage[hyperindex=true,
  pdfstartview=FitH,
  bookmarksnumbered=true,
  bookmarksopen=true,
  citecolor=blue,
  linkcolor=blue,
  colorlinks=true,
  pdfborder=001,
  unicode]{hyperref}

\allowdisplaybreaks

\newcommand{\D}{{\rm d}}
\newcommand{\ie}{\emph{i.e. }}

\renewcommand\){\right)}

\newcommand{\blue}[1]{\textcolor{blue}{#1}}

\renewcommand{\blue}[1]{\textcolor{black}{#1}}

\newcommand{\ind}[2]{^{#1}{}_{#2}}

\newcommand{\add}{\text{add}}
\newcommand{\fa}{\mathfrak{a}}
\newcommand{\fb}{\mathfrak{b}}

\newcommand{\re}{{\rm e}}
\newcommand{\pfrac}[2]{\frac{\partial #1}{\partial #2}}

\begin{document}

\title{General Relativistic Fluctuation Theorems}

\author{Yifan Cai}
\email{caiyifan@mail.nankai.edu.cn}
\affiliation{School of Physics, Nankai University, Tianjin 300071, China}

\author{Tao Wang}
\email{taowang@mail.nankai.edu.cn}
\affiliation{School of Physics, Nankai University, Tianjin 300071, China}

\author{Liu Zhao}\thanks{Corresponding author}
\email{lzhao@nankai.edu.cn}
\affiliation{School of Physics, Nankai University, Tianjin 300071, China}

\date{July 13, 2024}

\begin{abstract}
Using the recently proposed covariant framework 
of general relativistic stochastic mechanics and stochastic thermodynamics, 
we proved the detailed and integral fluctuation theorems in curved spacetime. 
The time-reversal transformation is described as 
a transformation from the perspective of future-directed observer to that of 
the corresponding past-directed observer, which enables us to 
maintain general covariance throughout the construction. 
\keywords{Langevin equation, fluctuation theorem, time-reversal symmetry, 
general relativity}
\end{abstract}

\maketitle

\emph{Introduction}. -- A fundamental problem in modern statistical physics is the 
emergence of macroscopic irreversibility in systems which have 
time-reversal symmetry (TRS) in the underlying microscopic description. 
The relevant researches can be traced back to Boltzmann, 
who employed a scattering model with TRS in deriving the H-theorem \cite{Boltzmann1970}, 
which states that the entropy of a macroscopic system cannot decrease
in the course of time. However, the H-theorem has been under debates and 
challenges ever since its birth. The most famous challenge is known 
as Loschmidt paradox \cite{ehrenfest1911conceptual,wu1975boltzmann}, which 
argues that, if an entropy-increasing process exists for a system, the underlying 
TRS should also permit a corresponding entropy-decreasing process. 
\blue{Nowadays, Loschmidt paradox is understood to be originated 
from the use of the molecular chaos hypothesis. After about one hundred and twenty years
since the birth of the H-theorem, a number of fluctuation theorems were proposed  
\cite{evans1994equilibrium, gallavotti1995dynamical, crooks1999entropy}, 
which provide an alternative {\em quantitative} description for the irreversibility 
of macroscopic systems by use of some equalities instead of the inequality 
presented by the H-theorem, and meanwhile attribute the origin of the irreversibility to 
the dissipative effects on the microscopic level.
}

In the context of stochastic mechanics, Sekimoto \cite{sekimoto1998langevin} 
utilized the overdamped Langevin equation to properly classify the energy exchange 
of a Brownian particle with the heat reservoir into trajectory heat and work, 
and thus establishes the first law of stochastic thermodynamics on the trajectory level. 
This allows for the construction of fluctuation theorems based on stochastic 
mechanics \cite{seifert2005entropy,imparato2006fluctuation,chernyak2006path,
ohkuma2007fluctuation,cai2024fluctuation}. 

The initial studies on fluctuation theorems are mostly carried out in 
non-relativistic theories. Since 2007, some attempts 
\cite{FINGERLE2007696,cleuren2008fluctuation,fei2019quantum,teixido2020first,
pal2020stochastic,pei2024special} in establishing 
fluctuation theorems in the special relativistic context appeared. 
Such attempts are important because 
relativity imposes stronger protection on the spacetime symmetry, making it harder 
to break the TRS. The purpose of the present work is to broaden the scope of the 
fluctuation theorem to encompass curved spacetime while maintaining general covariance. 
This is also important, because, on the one hand, gravity is a universal interaction,
it is desirable to see whether gravity has any impact on the origin of 
irreversibility --- another universal phenomenon that appear in the scope of 
macroscopic theories; on the other hand, the choice of time parameter is more subtle
in general relativistic theories than in special relativistic cases.

Conventionally, the time-reversal transformation (TRT) is merely described as a 
transformation of the time parameter, $t\mapsto -t$. However, if the time parameter 
is identified to be the coordinate time $x^0$, such a transformation will result 
in a lack of covariance. Our covariant framework \cite{cai2023relativistic,
cai2023relativistic2} for relativistic stochastic mechanics urges that the choice of 
time parameter should be closely linked with the observer. Consequently,  
the TRT needs be realized as a change from the perspective 
of a future-directed observer to that of a past-directed observer. 
In such a realization, the coordinate system is left intact.

\emph{Relativistic Langevin dynamics}.
-- Langevin equation describes the motion of a heavy particle, 
referred to as the Brownian particle, under the random disturbance of a 
heat reservoir. Conventionally, Langevin equation can be expressed in the form of 
Newton's second law, incorporating elements such as the random force, 
damping force, and various other external forces. For a relativistic 
particle of mass $m$ and charge $q$ moving in $(d+1)$-dimensional 
spacetime $\mathcal{M}$ with the metric $g_{\mu\nu}$, 
the Langevin equation employing the
particle's proper time $\tau$ as evolution parameter 
is referred to as LE$_\tau$ \cite{cai2023relativistic},  
\begin{align}\label{LEtau-em-1}
\D \tilde{x}_{\tau}^{\mu}&=\frac{\tilde{p}_{\tau}^{\mu}}{m}\D \tau,\\
\label{LEtau-em-2}
\D \tilde{p}_{\tau}^{\mu}&=\xi^\mu_\tau \D\tau
+\mathcal F_{\text{dp}}^\mu\D \tau
-\frac{1}{m}\varGamma\ind{\mu}{\alpha\beta}\tilde{p}_{\tau}^{\alpha}
\tilde{p}_{\tau}^{\beta}\D \tau
+\mathcal F_{\text{em}}^\mu \D\tau,
\end{align}
where $\xi^\mu_\tau$ is the random force, $\mathcal F_{\text{dp}}^\mu:=
\mathcal{K}^{\mu\nu}U_\nu$ is the damping force
in which the damping coefficient tensor $\mathcal{K}^{\mu\nu}$ 
obeys $\mathcal{K}^{\mu}{}_{\nu}p^\nu=0$, $\displaystyle\mathcal F_{\text{em}}^\mu:=
\frac{q}{m}F^{\mu}{}_{\nu}\tilde{p}^{\nu}_{\tau}$ is the electromagnetic force,
and $\varGamma\ind{\mu}{\alpha\beta}$ denotes the Christoffel connection 
associated with $g_{\mu\nu}$.
Tilded symbols such as $\tilde x_\tau$ and $\tilde p_\tau$ represent random 
variables, and the un-tilded ones represent their realization. 
The random force is consisted of a term encoding the Stratonovich 
coupling between the stochastic amplitudes $\mathcal{R}\ind{\mu}{\fa}$ 
(which transforms as a vector field for each fixed $\fa$) with a 
set of $d$ independent Wiener process $\D\tilde w_\tau^\fa$ obeying 
the probability distribution
\begin{align}
\Pr[\D\tilde w^\fa_\tau=\D w^\fa]=\frac{1}{(2\pi\D \tau)^{d/2}}
\exp\left[-\frac{\delta_{\fa\fb}\D w^\fa\D w^\fb}{2\D \tau}\right]
\end{align}
of variance $\D\tau$, 
together with a term incorporating an additional stochastic force
$\displaystyle\mathcal{F}^\mu_\add:=
\frac{\delta^{\fa\fb}}{2}\mathcal R\ind{\mu}{\fa}\nabla^{(h)}_i \mathcal R\ind{i}{\fb}$,
\begin{align}\label{eq:random-force-tau}
\xi_\tau^\mu:=\mathcal R\ind{\mu}{\fa}\circ_S\D\tilde w^\fa_\tau/\D\tau
+\mathcal{F}^\mu_\add,
\end{align}
where $\nabla^{(h)}$ is the covariant derivative on the mass shell
and $\mathcal R\ind{\mu}{\fa}$ stands for the stochastic amplitude which 
may depend on $\tilde{x}_{\tau}^{\mu}$ and $\tilde{p}_{\tau}^{\mu}$ and
transform as a vector field
for each fixed $\fa$.
The Stratonovich coupling $\circ_S$ maintains the chain rule in stochastic calculus, 
making the general covariance of LE$_\tau$ self-evident.

The energy of a relativistic particle measured by the observer with proper velocity 
$Z^\mu$ is defined as
\begin{align}
E(x,p):=-Z_{\mu}(x)p^{\mu}.
\end{align}
As the microstate of the Brownian particle evolves in a stochastic process, 
the energy of the Brownian particle also becomes a stochastic process 
$\tilde E_\tau:=E(\tilde x_\tau,\tilde p_\tau)$. The chain rule 
implies that
\begin{align}\label{eq:dE}
\D\tilde E_\tau
&=\pfrac{E}{x^\mu}\D\tilde x^\mu_\tau+\pfrac{E}{p^\mu}\D\tilde p^\mu_\tau\notag\\
&=-Z_\mu\left[\xi_\tau^\mu+\mathcal F_{\text{dp}}^\mu \right]\D\tau
-\frac{\tilde p_\tau^\mu \tilde p_\tau^\nu}{m}\nabla_\nu Z_\mu \D\tau 
-Z_\mu \mathcal F_{\text{em}}^\mu\D\tau.
\end{align}
The second and last terms in eq.~\eqref{eq:dE} are recognized to be the 
gravitational \cite{liu2021work} and electromagnetic works,
\begin{align}
\D\tilde{\mathcal P}_\tau
:=-\frac{\tilde p_\tau^\mu \tilde p_\tau^\nu}{m}\nabla_\nu Z_\mu \D\tau, \quad
\D\tilde{\mathcal W}_\tau:=-Z_\mu\mathcal{F}^\mu_{\text{em}}\D\tau,
\end{align}
and the first term incorporates the energy change caused by the impact of the heat reservoir,  
which is by definition the heat received by the Brownian particle, 
\begin{align}
\D\tilde Q_\tau:=-Z_\mu\left[\xi_\tau^\mu+\mathcal F_{\text{dp}}^\mu \right]\D\tau.
\end{align}
Therefore, eq.~\eqref{eq:dE} becomes precisely the first law of 
general relativistic stochastic thermodynamics,
\begin{align}\label{first-law}
\D\tilde E_\tau=\D\tilde{\mathcal Q}_\tau+\D\tilde{\mathcal P}_\tau
+\D\tilde{\mathcal W}_\tau.
\end{align}

\emph{Time-reversal symmetry}.
-- Given a worldline $x_\tau$ for a massive particle in the spacetime $\mathcal{M}$, 
its tangent vector $p^\mu_\tau$ can be either {\em aligned} with the proper velocity 
$Z_\mu$ of some chosen observer Alice, \ie $p^\mu_\tau Z_\mu <0$, or {\em opposite} to the 
proper velocity $C_\mu$ of some other observer, Carol, \ie $p^\mu_\tau C_\mu >0$. 
We will refer to Alice as {\em future-directed} and to Carol as {\em past-directed}. 
To be more specific, we fix Carol to be the TRT image of Alice, \ie $C_\mu=-Z_\mu$. 
If Alice perceives the worldline evolving from $t_I$ to $t_F$, then Carol 
will perceive the worldline evolving from $t_F$ to $t_I$.
This leads to a change in the sign of the time derivatives, 
{\em e.g.} $\D/\D\tau \rightarrow -\D/\D\tau$, 
and consequently the momentum $p$ needs to reverse its sign. Notice however that 
$\D\tau$ resp. $\D t$ denote the size of infinitesimal temporal steps, both remain 
unchanged under the TRT.

The mass shell bundle $\Gamma_m:=\{(x,p)\in T\mathcal{M}|p_\mu p^\mu=-m^2 \}$ 
can be separated into two non-path-connected 
regions, \ie the future mass shell bundle
\begin{align}
\Gamma^+_m:=\{(x,p)\in\Gamma_m|Z_\mu p^\mu<0\},
\end{align}
and the past mass shell bundle
\begin{align}
\Gamma^-_m:=\{(x,p)\in\Gamma_m|Z_\mu p^\mu>0\},
\end{align}
both are defined relative to Alice. 
Clearly, there is a homeomorphism between $\Gamma^+_m$  and $\Gamma^-_m$:
\begin{align}
I: \Gamma^+_m\to \Gamma^-_m,\qquad (x,p)\mapsto (x,-p).
\end{align}
This homeomorphism is the mathematical realization of the TRT. 
The phase trajectory $Y_t=(y_t,k_t)$ 
of the Brownian particle is the uplift of the worldline into 
$\Gamma^+_m$, and is referred to as the {\em forward trajectory}, 
while the TRT image of the phase trajectory $Y^-_t=I(Y_{t_I+t_F-t})$ 
is the uplift of the worldline 
into $\Gamma^-_m$, and is known as the {\em reversed trajectory}. 

Now consider a macroscopic system consisting of a great number of 
massive charged particles which can be classified into several species. 
The electromagnetic field is produced by the charges carried by the particles
and the geometry of the $(d+1)$-dimensional spacetime is determined by the 
masses and charges within the system.
The phase trajectories of these particles obey the following equations, 
\begin{align}\label{5.33}
\frac{\D x^\mu_s}{\D \tau_s}&=\frac{p^\mu_s}{m_s},\\
\label{5.34}
\frac{\D p^\mu_s}{\D \tau_s}&=\frac{q_s}{m_s}F^{\mu}{}_{\nu}p^\nu_s
-\frac{1}{m_s}\varGamma\ind{\mu}{\alpha\beta}p^\alpha_s p^\beta_s,
\end{align}
where the suffices $s$ indicate different species. 
Assuming that each species of particles obeys a distinct 
TRT invariant distribution $\Phi_s(x_s,p_s)$,
the electric current and the energy-momentum tensor 
contributed by the particles can be written as
\begin{align}\label{electric-current}
J^\mu&=\sum_s \int \varpi_s q_s p^\mu_s\Phi_s,\\
\label{tensor-particles}
\qquad 
T^{\mu\nu}_{\text{pa}}&=\sum_s \int \varpi_s p^\mu_s p^\nu_s\Phi_s,
\end{align}
where $\varpi_s:=(\sqrt{g}/(p_s)_0) \D p^1_s\wedge...\wedge\D p^d_s$ is the invariant 
volume element on the mass shell of the $s$-species. The Maxwell equation 
and the energy-momentum tensor produced by the electromagnetic field 
are given by
\begin{align}\label{maxwell}
\nabla_\nu F^{\mu\nu}=J^\mu, \quad 
T^{\mu\nu}_{\text{em}}= F^{\mu \rho} F\ind{\nu}{\rho}
-\frac{1}{4}g^{\mu\nu}F^{\rho\sigma}F_{\rho\sigma},
\end{align}
and finally, the Einstein equation that determines the spacetime geometry 
reads 
\begin{align}\label{5.37}
R^{\mu\nu}-\frac{1}{2}g^{\mu\nu} R=8\pi G\(T^{\mu\nu}_{\text{pa}}+T^{\mu\nu}_{\text{em}}\).
\end{align}

As we shift the perspective from Alice to Carol, there exist multiple equivalent 
conventions \cite{arntzenius2009time,cai2024fluctuation2} regarding the 
TRT of the electromagnetic field and the charge. We adopt Feynman's
convention \cite{feynman1965nobel}:
\begin{align}
q\mapsto -q, \quad F^{\mu\nu}\mapsto F^{\mu\nu}.
\end{align}
It becomes straightforward to verify that all equations of 
motion for the above system, \ie eqs.~\eqref{5.33}-\eqref{5.37}, are  
invariant under the TRT. Therefore, 
in the presence of TRS, our designation of Alice as future-directed observer 
has no microscopic meaning: one cannot distinguish Alice from Carol purely from
the mechanical description.

Stochastic mechanics, in essence, serves as an effective theory for  
complicated mechanical systems within specific spatial and temporal scales 
\cite{ford1965statistical,mori1965transport,zwanzig1973nonlinear}. Consider a 
single heavy particle in the above system, and assuming that the remaining 
particles constitute a heat reservoir which have already attained thermal equilibrium.
Due to the presence of gravity, the equilibrium state 
is {\em not} a state with uniform temperature, but rather a state of 
a fluid with the temperature obeying \cite{Eckart}
\begin{align}
\nabla_{\mu} T_{\rm B} + {T_{\rm B}} U^{\nu} \nabla_{\nu} U_{\mu} = 0,
\label{TE}
\end{align}
where $U^\mu$ denotes the proper velocity of the fluid element. 
The non-uniform distribution of the temperature described by eq.~\eqref{TE}
is known as Tolman-Ehrenfest effect. Please be reminded that, in relativistic context, 
the temperature is observer-dependent. The particular temperature $T_{\rm B}$
appeared above is the one perceived by the observer comoving with the heat reservoir,
Bob, as indicated by the suffix $\rm B$.

The electromagnetic interaction exerted on the heavy particle can be separated  
into two components: the coarse-grained averaging effects at larger spatial 
and temporal scales, and the stochastic approximations at smaller scales. 
The latter encompasses both the random force and damping force. As a result, 
eqs.~\eqref{5.33}-\eqref{5.34} can be approximated by LE$_\tau$ 
\cite{cai2023relativistic}, rendering this heavy particle as the Brownian particle. 
It will be clear shortly that the above coarse-grained picture breaks the TRS.

\emph{Fluctuation theorem}.
-- Selecting an integral curve of Alice, we can interpret its arc length as 
the proper time $t$ of Alice, which can be further extended as a scalar field 
on $\mathcal M$ \cite{cai2023relativistic}. The proper time of Alice and that of 
the Brownian particle are connected via the following relation:
\begin{align}
\D t=\frac{p^\mu}{m}\partial_\mu t\D\tau=\gamma(x,p)\D\tau,
\end{align}
where $\gamma(x,p)$ is the local Lorentz factor arising from the relative motion between 
Alice and the particle. 
Since the microstate $(x,p)$ of the Brownian particle is random, 
the infinitesimal increment of $\tau$ also becomes random from Alice's 
perspective. Consequently, for Alice, LE$_\tau$ needs to be reparameterized 
using the deterministic parameter $t$ in place of the random parameter $\tau$, 
yielding a new form of Langevin equation known as LE$_t$ 
\cite{cai2023relativistic}, 
\begin{align}
\D\tilde y_t^\mu&=\frac{\tilde k^\mu_t}{m}\gamma^{-1}\D t,\\
\D\tilde k_t^\mu&=\hat \xi_t^\mu \gamma^{-1}\D t
+\mathcal{F}^\mu_{\text{dp}}\gamma^{-1}\D t\notag\\
&+\mathcal{F}^\mu_{\text{em}}\gamma^{-1}\D t
-\frac{1}{m}\varGamma\ind{\mu}{\alpha\beta}\tilde k^\alpha_t\tilde k^\beta_t \gamma^{-1}\D t,
\end{align}
where $\tilde Y_t=(\tilde y_t,\tilde k_t)=(\tilde x_{\tilde{\tau}_t},
\tilde p_{\tilde{\tau}_t})$, and 
\begin{align}
\hat\xi^\mu_t:=\gamma^{1/2}\mathcal{R}\ind{\mu}{\fa}\circ_S\D\tilde W_t^\fa/\D t
+\mathcal{F}^\mu_\add-\frac{1}{2}\mathcal{D}^{\mu i}\gamma^{1/2}\nabla^{(h)}_i\gamma^{-1/2},
\end{align}
in which $\D\tilde W_t^\fa$ represents a Wiener process of variance $\D t$ for each
fixed $\fa$. 
It is evident that LE$_t$ is observer-dependent, but still manifestly general covariant. 
The stochastic process $\tilde Y_{[t]}$ reparameterized by Alice is referred to 
as the \emph{forward process}, while the stochastic process $\tilde Y^-_{[t]}$
reparameterized by Carol is referred to as the \emph{reversed process}. 
It is worth noticing that {\em process} and {\em trajectory} are different concepts, 
the latter is a concrete realization of the former. 
In particular, the initial state of the reversed process and final state of the 
forward process are related via 
\begin{align}\label{relation-process}
\tilde Y^-_I=I(\tilde Y_F),
\end{align}
and there {\em need not} be any relationship between other states from the forward and 
reversed processes.

Contrary to conventional mechanical equations, Langevin equation lacks the 
capacity to determine whether a trajectory qualifies as its solution. Instead, 
the best one can do is to determine the probability of a given trajectory.  
Correspondingly, the TRS breaking manifests as the difference in the 
probabilities of the forward and reversed trajectories,
\begin{align}
\Pr[\tilde Y_{[t]}=Y_{[t]}]\neq \Pr[\tilde Y_{[t]}^-=Y_{[t]}^-].
\end{align}

In order to quantify to what extent the TRS is broken in the general 
relativistic Langevin dynamics described by LE$_t$, let us turn to the construction of 
the fluctuation theorem. The stochastic process governed by LE$_t$
is a Markov process. Consequently, the trajectory probability can be expressed 
as the product of the initial and conditional probabilities:
\begin{align}
\Pr[\tilde Y_{[t]}=Y_{[t]}]=\Pr[\tilde Y_{[t]}=Y_{[t]}|\tilde Y_{I}= Y_I]
\Pr[\tilde Y_{I}=Y_I],
\end{align}
and the conditional probability can be further decomposed into the product of 
the transition probabilities for a sequence of intermediate steps \cite{cai2023relativistic2}. 
The initial probability $\Pr[\tilde Y_{I}=Y_I]$ is related to 
the one particle distribution $\varphi(Y_I)$ of the relativistic Brownian particle 
via
\begin{align}
\Pr[\tilde Y_{I}=Y_I]=\gamma\lambda^{-1}\varphi(Y_I),\quad \lambda:=|\partial_\mu t|,
\end{align}
where $\varphi(Y_I)$ obeys the reduced Fokker-Planck equation 
\cite{cai2023relativistic2}, but is {\em not} a probability distribution by itself.
The entropy density of the Brownian particle 
should be defined as \cite{wang2023general}
\begin{align}\label{entropy-Brownian}
S:=-\ln\varphi,
\end{align}
the difference of which at the final and the initial states of a trajectory is 
defined to be the {\em trajectory entropy production}.
Furthermore, the ratio of the conditional probabilities of the forward and reversed
trajectories is given by \cite{cai2024fluctuation2}:
\begin{align}\label{rate-condition-probability}
&\frac{\Pr[\tilde Y_{[t]}=Y_{[t]}|\tilde Y_{I}= Y_I]}
{\Pr[\tilde Y_{[t]}^-=Y^-_{[t]}|\tilde Y^-_{I}= Y^-_I]}\notag\\
=&\frac{(\lambda\gamma^{-1})|_{Y_I}}{(\lambda\gamma^{-1})|_{Y_F}}
\exp\left[\int^{t_F}_{t_I}\gamma^{-1}\D t\frac{1}{T_{\rm B}}
(ma^\mu-\mathcal{F}^\mu_{\text{em}})U_\mu \right],
\end{align}
where $a^\mu=p^\nu\nabla_\nu p^\mu/m^2$ is the proper acceleration of the Brownian particle. 
From the perspective of Bob, the complete differential of the energy is:
\begin{align}
\D E=-\frac{p^\nu}{m}\nabla_\nu(p^\mu U_\mu)\D \tau=-ma^\mu U_\mu\D\tau+\D\mathcal{P}.
\end{align}
Comparing with eq.~\eqref{first-law}, it can be deduced that the integral over 
the exponent in eq.~\eqref{rate-condition-probability} is equivalent to the entropy 
increase of the heat reservoir, thanks to the relativistic Clausius equality at 
the trajectory level:
\begin{align}\label{Clausius}
\int\gamma^{-1}\D t\frac{1}{T_{\rm B}}(m a^\mu-\mathcal{F}^\mu_{\text{em}})U_\mu
=-\int\frac{\D \mathcal Q}{T_{\rm B}}=\Delta S_R.
\end{align}
The relativistic Clausius equality at the ensemble level has been discussed 
in our previous work \cite{wang2023general}.
Although the relativistic Clausius equalities are only valid from the perspective of Bob, 
the amount of the entropy increase is observer-independent.
Therefore, the ratio of the probabilities of the forward and reversed trajectories 
can be deduced from eq.~\eqref{entropy-Brownian} and 
eq.~\eqref{rate-condition-probability} ,
\begin{align}\label{fluctuation-theorem-detail1}
\frac{\Pr[\tilde Y_{[t]}=Y_{[t]}]}{\Pr[\tilde Y^-_{[t]}=Y^-_{[t]}]}
=&\frac{\Pr[\tilde Y_{[t]}=Y_{[t]}|\tilde Y_{I}= Y_I]}
{\Pr[\tilde Y_{[t]}^-=Y_{[t]}^-|\tilde Y_{I}^-= Y_I^-]}
\frac{\Pr[\tilde Y_I=Y_I]}{\Pr[\tilde Y_I^-=Y_I^-]}\notag\\
=&\re^{\Delta S_R}\frac{(\lambda\gamma^{-1})|_{Y_I}}{(\lambda\gamma^{-1})|_{Y_F}}
\frac{\Pr[\tilde Y_I=Y_I]}{\Pr[\tilde Y_F=Y_F]}\notag\\
=&\re^{\Delta S_R+\Delta S}.
\end{align}

Let $\Sigma_{Y_{[t]}}=\Delta S_R+\Delta S$ be the total entropy production contributed 
by the trajectory $Y_{[t]}$, eq.~\eqref{fluctuation-theorem-detail1} 
can be rewritten as 
\begin{align}\label{ftd}
\frac{\Pr[\tilde Y_{[t]}=Y_{[t]}]}{\Pr[\tilde Y^-_{[t]}=Y^-_{[t]}]}=\re^{\Sigma_{Y_{[t]}}}.
\end{align}
This is the standard form of fluctuation theorem on the trajectory level.
Furthermore, by integration over the space of trajectories
and employing Jensen's inequality, we arrive at the integral fluctuation theorem:
\begin{align}\label{fti}
\re^{-\left\langle \Sigma_{\tilde Y_{[t]}}\right\rangle}&\leq 
\left\langle \re^{-\Sigma_{\tilde Y_{[t]}}} \right\rangle\notag\\
&=\int\mathcal{D}[Y_{[t]}] \re^{-\Sigma_{\tilde Y_{[t]}}}
\Pr[\tilde Y_{[t]}=Y_{[t]}]\notag\\
&=\int\mathcal{D}[Y_{[t]}]\Pr[\tilde Y^-_{[t]}=Y^-_{[t]}]\notag\\
&=\int\mathcal{D}[Y^-_{[t]}]\Pr[\tilde Y^-_{[t]}=Y^-_{[t]}]=1,
\end{align}
which means the statistical expectation value of the total entropy production 
must be no-negative, \ie $\left\langle \Sigma_{\tilde Y_{[t]}}\right\rangle\geq 0$. 
For most trajectories, the total entropy production $\Sigma_{Y_{[t]}}$ is nonzero, 
indicating distinct probabilities for forward and reversed trajectories. 
The fluctuation theorem thus provides a quantitative link between TRS breaking 
and entropy production.

\emph{Conclusion.}
-- Based on the covariant framework of general relativistic stochastic mechanics,
we formulated the first law of stochastic thermodynamics in curved spacetime. 
By properly addressing the description for the TRS and its breaking, 
the corresponding fluctuation theorem is also proved in both the differential 
and integral forms. The observers and their behaviors under the TRT 
play a key role in our construction. The fact that the final fluctuation theorems
\eqref{ftd} and \eqref{fti} take the same form as in non-relativistic context
\cite{seifert2005entropy} is expected, because both the trajectory probability 
and the total entropy production are neither coordinate dependent nor 
observer dependent. 


\blue{The proof of the fluctuation theorems presented in this work 
extends the range of applicability of the fluctuation theorems to 
the cases involving strong gravity and heat reservoir with a non-uniform temperature. 
We expect that such extension may be found useful in certain astrophysical processes. 
We hope to come back on this point in future works.}

\emph{Acknowledhement} -- This work is supported by the National Natural Science 
Foundation of China under the grant No. 12275138.

\emph{Conflict of interests declaration:} -- The authors declare no 
known conflict of interests.

\providecommand{\href}[2]{#2}\begingroup
\footnotesize\itemsep=0pt
\providecommand{\epr}[2][]{\href{http://arxiv.org/abs/#2}{arXiv:#2}}

\end{document}